\documentclass[aps,prb,twocolumn,showpacs,letterpaper,superscriptaddress]{revtex4-1}

\usepackage[colorlinks=true, citecolor=blue, linkcolor=blue, urlcolor=blue]{hyperref}
\usepackage{graphicx,dcolumn,longtable,epsfig}
\usepackage[usenames]{color}
\usepackage{amssymb}
\usepackage{amsmath}
\usepackage{bm}
\usepackage{footnote}
\usepackage{float}
\usepackage{subfigure}
\usepackage{color}

\usepackage{ulem}
\usepackage[T1]{fontenc}

\newcommand{\be}{\begin{equation}}
\newcommand{\ee}{\end{equation}}

\def\bea{\begin{eqnarray}}
\def\eea{\end{eqnarray}}

\begin{document}

\title{Effect of dispersive optical phonons on the properties of Bond Su-Schrieffer-Heeger polaron}

\author{Chao Zhang}
\email{zhangchao1986sdu@gmail.com}
\affiliation{Department of Modern Physics, University of Science and Technology of China, Hefei, Anhui 230026, China}
\affiliation{Hefei National Laboratory, University of Science and Technology of China, Hefei, Anhui 230088, China}

\begin{abstract}

%We use a newly developed Quantum Monte Carlo method based on the path-integral formulation of the particle sector and the real-space diagrammatic technique of the phonon sector to investigate the impact of a finite dispersion of the optical phonon mode on the properties of the bond Su-Schrieffer-Heeger polaron in two dimensions. Our study focuses on the effect of positive bandwidth on the bond polaron's effective mass, ground state energy, and $Z$ factor, comparing these properties with those of the dispersionless case. At the same electron-phonon coupling strength, in the absence of phonon dispersion, we observe that the effective mass increases as the phonon frequency decreases, indicating a heavier polaron in the deep adiabatic regime. While in the dispersive case, we find that the effective mass increases as the phonon bandwidth increases. Furthermore, we observe a crossover from a light polaron state to a heavy polaron state in the deep adiabatic regime for both the dispersionless and dispersive case. 

We use a newly developed Quantum Monte Carlo method to investigate the impact of finite dispersion of the optical phonon mode on the properties of the bond Su-Schrieffer-Heeger (SSH) polaron in two dimensions. We compare the properties of the bond polaron, such as effective mass, ground state energy, and $Z$ factor, with and without positive phonon bandwidth. Our findings reveal that when we exclude phonon dispersion, at the same electron-phonon coupling strength, the effective mass increases as the phonon frequency decreases, indicating a heavier polaron in the deep adiabatic regime. However, in the dispersive case, we observe that the effective mass increases as the phonon bandwidth increases. Moreover, we notice a crossover from a light polaron state to a heavy polaron state in the deep adiabatic regime for both the dispersionless and dispersive cases.

\end{abstract}

\maketitle

%%%%%%%%%%%%%%%%%%%%%%%%%%%%%%%%%%%%%%%%%%%%%%%%%%%%%%%%%%%%%%%%%%%%%%%%%
\section{Introduction}

%Polaron problem keeps attracting attention in decants and it asks what happens when a particle coupled to an environment and what is the property of the resulting object, which is called a polaron. Depending on the properties of the particle, the environment, and their coupling, different types of polarons can arise, such as electron-phonon polarons~\cite{Landau33,Frohlich50,Feynman:1955du, Schultz:1959el, Holstein59, Alexandrov:1999fy, Holstein2000}, spin-polarons~\cite{BR70,Nagaev:1974jb, Mott:2006fa}, Fermi-polarons~\cite{Bulgac:2007bl, Lobo:2006dc, Prokofev:2008jz, Prokofev:2008it}, protons in neutron rich matter \cite{protons}. Among these, the electron-phonon polaron is particularly important because it is associated with the mechanism of high-temperature superconductivity in the dilute-density regime. In the low-density limit, the electron-phonon interaction can combine two polarons into a single bipolaron, forming a Bose-Einstein condensate-like superconductor. However, for such a superconductor to occur, a bipolaron with a light effective mass and a strong phonon-mediated pairing potential is required.

The polaron problem continues to draw significant attention in condensed matter physics. It studies the behavior of a particle when coupled to its environment. Depending on the characteristics of the particle, the environment, and their coupling, various types of polarons can arise, including polarons with electron-phonon interaction~\cite{Landau33,Frohlich50,Feynman:1955du, Schultz:1959el, Holstein59, Alexandrov:1999fy, Holstein2000}, spin-polarons~\cite{BR70,Nagaev:1974jb, Mott:2006fa}, Fermi-polarons~\cite{Bulgac:2007bl, Lobo:2006dc, Prokofev:2008jz, Prokofev:2008it}, and protons in neutron-rich matter \cite{protons}. Of particular significance is the polaron with electron-phonon interaction, as it plays a crucial role in understanding the mechanism of high-temperature superconductivity in the dilute-density regime. In the low-density limit, the electron-phonon interaction can bind two polarons together to form a single bipolaron, forming of a Bose-Einstein condensate-like superconductor. However, for such a superconducting state to manifest, it is essential to have a bipolaron with a light effective mass and a robust phonon-mediated pairing potential.

%Previous studies has shown that in the Holstein model where the electron-phonon coupling affects on the density of the electrons, the effective mass of polaron and bipolaron increase exponentially at strong electron-phonon coupling strengths~\cite{PhysRevB.55.R8634, PhysRevLett.105.266605, PhysRevLett.84.3153, PhysRevB.69.245111}. However, when the electron-phonon coupling affects the hopping of electrons, which is known as the bond Su-Schrieffer-Heeger polaron (SSH), the situation becomes different~\cite{PhysRevB.56.4484, PhysRevB.69.174301}. Recently, there has been a surge of interest in studying bond SSH polarons and bipolarons, since in the absence of phonon dispersion, the effective mass is not exponentially large at strong coupling regimes, resulting in a light polaron~\cite{PhysRevLett.105.266605, PhysRevB.104.035143, PhysRevB.104.L140307}. Moreover, when two polarons form a bound state, the effective mass of the resulting bipolaron is still light enough~\cite{PhysRevLett.105.266605} and the size of the bipolaron is small, which leads to the possibility of high-temperature superconductivity~\cite{PhysRevX.13.011010,Johnarxiv2022}. In all of these cases, only phonons with the Einstein mode are considered. One way to extend the bond SSH model is to introduce dispersion among localized (Einstein) phonons. However, there are very few attempts in the literature to study polaron/bipolaron models with dispersive phonons due to the scarcity of numerical techniques suitable for this type of analysis even in the most-well known Holstein model.

Previous studies have demonstrated that in the Holstein model, where the electron-phonon coupling influences the electron density, both the effective mass of the polaron and the bipolaron exhibit an exponential increase under strong electron-phonon coupling strengths~\cite{PhysRevB.55.R8634, PhysRevLett.105.266605, PhysRevLett.84.3153, PhysRevB.69.245111}. However, when the electron-phonon coupling affects the electron hopping, as seen in the bond Su-Schrieffer-Heeger (SSH) polaron, the situation takes a different turn~\cite{PhysRevB.56.4484, PhysRevB.69.174301}. Recently, there has been a notable surge in interest in studying bond SSH polarons and bipolarons, particularly when phonon dispersion is absent. In such cases, the effective mass is not exponentially large even at strong coupling regimes, resulting in the formation of light polarons~\cite{PhysRevLett.105.266605, PhysRevB.104.035143, PhysRevB.104.L140307}. Remarkably, when two polarons form a bound state, the resulting bipolaron maintains a sufficiently low effective mass~\cite{PhysRevLett.105.266605}, and its size remains small, indicating the potential for high-temperature superconductivity~\cite{PhysRevX.13.011010,Johnarxiv2022}. It's important to note that in all of the mentioned cases, only phonons with the Einstein mode are considered. Extending the bond SSH model to include dispersion among localized (Einstein) phonons opens up new possibilities for investigation. However, such an extension presents challenges due to the scarcity of numerical techniques suitable for analyzing polaron/bipolaron models with dispersive phonons, even in the well-known Holstein model. As a result, there have been limited attempts in the literature to study such systems with dispersive phonons~\cite{PhysRevB.106.155129, PhysRevB.106.174303}.

In the dilute density limit, in the Holstein model, the phonon degrees of freedom dress the electrons, giving rise to polaron and bipolaron formation. Previous studies~\cite{PhysRevB.88.060301, PhysRevB.103.054304} have investigated the influence of the dispersion among optical phonons on the polaron's effective mass in one dimension. However, Ref~\cite{PhysRevB.88.060301} is limited to one dimensional which is not physical, and the adiabatic regime $\omega_0/t <1$ has not been explored due to the computational hard. At higher densities, the phonons mediate collective superconducting and charge-density wave phases. %Recently, a study has demonstrated the significant influence of phonon dispersion on the formation of charge-density-wave order in a system with finite-electron density using the quantum Monte Carlo technique~\cite{PhysRevLett.120.187003}. However, the effect of finite dispersion of the optical phonon mode on the properties of the bond SSH model, particularly the effective mass, is still unknown.
A recent study employing the quantum Monte Carlo technique has demonstrated the significant influence of phonon dispersion on the formation of charge-density-wave order in a system with finite electron density~\cite{PhysRevLett.120.187003}. However, the specific impact of finite dispersion of the optical phonon mode on the properties of the bond SSH model, particularly with regard to the effective mass, remains unexplored.

%In this paper, we investigate the effects of a finite dispersion of the optical phonon mode on the properties of the bond SSH polaron in two dimensions using a newly developed Quantum Monte Carlo method. Our approach is based on the path-integral formulation of the particle sector combined with real space diagrammatic techniques for the phonon sector~\cite{PhysRevB.105.L020501}. We consider only the positive phonon bandwidth due to the `sign' problem that arises with negative phonon bandwidth in this method. To the best of our knowledge, our theory is the first quantitative effect to (i) demonstrate, using an unbiased approach, the properties of bond polaron with the dispersive phonons in two dimensions with phonon bandwidth $W$ smaller, equal, and larger than the phonon frequency $\omega_0$ and (ii) study the properties in the adiabatic regime as low as $\omega_0/t=0.3$.

In this paper, we utilize a newly developed Quantum Monte Carlo method to explore the influence of finite dispersion in the optical phonon mode on the properties of the bond SSH polaron in two dimensions. Our methodology combines the path-integral formulation for the particle sector with real-space diagrammatic techniques for the phonon sector~\cite{PhysRevB.105.L020501}. Due to the `sign' problem associated with negative phonon bandwidth in this approach, we focus solely on the positive phonon bandwidth. To the best of our knowledge, our study is the first to quantitatively investigate the properties of the bond polaron with dispersive phonons in two dimensions, encompassing phonon bandwidths $W$ smaller, equal, and larger than the phonon frequency $\omega_0$. Additionally, we explore properties in the adiabatic regime down to $\omega_0/t=0.3$. The rest of the paper is organized as follows. In Sec.~\ref{sec:sec2}, we present the Hamiltonian of the bond SSH polaron. In Sec.~\ref{sec:sec3}, we introduce how to extract the properties from the Green's function. In Sec.~\ref{sec:sec4}, we discuss the results and Sec.~\ref{sec:sec5} concludes the paper. 

%Just be careful with the sign of the phonon propagator when the phonon hopping amplitude changes sign. I think is will become negative for displacements ix+iy = odd integer.

\section{Hamiltonian}
\label{sec:sec2}

We consider a bond SSH electron-phonon coupling on a simple two-dimensional square lattice. In this model, the electronic hopping between two sites is modulated by a single oscillator centered on the bond connecting the two sites. The model is described by the Hamiltonian~\cite{PhysRevLett.42.1698, PhysRevLett.25.919, PhysRevB.5.932, PhysRevB.5.941}:
\begin{equation}
\begin{split}
H_1 =& -t \sum_{\langle i,j \rangle, \sigma}(c_{j,\sigma}^{\dagger}c_{i, \sigma} + h. c.) +\omega_0 \sum_{\langle i, j \rangle} \bigg{(}b^{\dagger}_{i, j} b_{i, j } +\frac{1}{2} \bigg{)} \\
& + g  \sum_{\langle i, j \rangle, \sigma} \bigg{(} c_{j, \sigma}^{\dagger} c_{i, \sigma} + h. c.  \bigg{)} \bigg{(}b_{i, j}^{\dagger} + b_{i, j} \bigg{)}  ,
\label{H1}
\end{split}
\end{equation}
where $\langle i, j \rangle$ denotes the nearest-neighbor sites. $b_i$ is the optical phonon annihilation operator on site $i$ and $c_{i,\sigma}$ is the annihilation operator for electron on site $i$ with spin $\sigma \in \{ \uparrow, \downarrow \}$, $t$ is the electron hopping amplitude between the nearest-neighbor sites (we use it as the unit of energy), and $g$ is the strength of the electron-phonon coupling of the hopping-displacement type. $\omega_0$ is the local phonon frequency. 

We generalize Eq.(1) to $H=H_1+H_2$, to include a coupling strength $t_{ph}$ between nearest-neighbor bonds for the phonons, with
\begin{equation}
H_2 =-t_{ph} \sum_{\langle \langle i, j\rangle , \langle i', j' \rangle \rangle} \bigg{(} b^{\dagger}_{i, j} b_{i', j' } + h.c.    \bigg{)}  .
\label{H2}
\end{equation}
Here $\langle \langle i, j \rangle , \langle i', j' \rangle \rangle$ denotes nearest-neighbor bonds. $t_{ph}$ is the hopping amplitude for phonons between nearest-neighbor bonds, and it can be positive or negative. %For $t_{ph} < 0(> 0)$, we refer to the phonons as having an upward (downward) dispersion relation. Physically, the minus sign is the more natural one and representing the forces on atoms depend on their relative displacement. The positive sign yields a mode with a downward bending momentum $0$ to $\pi$, the more typical behavior for high-frequency optical modes. 
The sign of the phonon propagator changes when the phonon hopping amplitude $t_{ph}<0$, causing the `sign' problem. So here, we only consider $t_{ph} > 0$. The inclusion of the nearest-neighbor hopping of the phonons $t_{ph} > 0$ leads to a finite phonon bandwidth $W=8 t_{ph}$. When considering the dispersive case, some of the phonons are softer and the phonon frequency is lowered to $\omega_L=\omega_0-W/2$.
%In this case the lowest frequency goes down as  w_L = w0 -W_{ph}/2 and the effect is suppressed mass relative to dispersion-less w0 because we some of the phonons are softer. 

The properties of the bond SSH polaron %for Einstein phonons ($W=0$) 
is controlled by two dimensionless parameters: (i) the effective coupling 
\begin{equation}
\lambda=\frac{g^2}{2 t \omega} ,
\label{lambda}
\end{equation}
here, $\omega$ is defined as $\omega=\omega_0$ with $\omega_0$ the local phonon frequency for the dispersionless case, and $\omega=\omega_L$ for the dispersive case. %{\cz rewrite?}%with strong coupling regime corresponding to $\lambda \ge 0.25$ in this definition. 
(ii) The adiabaticity ratio $\omega_0/t$. Here, in this paper, we work on the adiabatic regime $\omega_0/t \le 1.0$ where the phonon degree of freedom is considered comparable or slow with respect to the electron motion. 

\section{Green's function}
\label{sec:sec3}
In the following, we study the effect of the positive phonon bandwidth $W$ at a certain phonon frequency $\omega_0$ on the properties of the bond polaron: the ground state energy, the effective mass as well as the $Z$ factor. The polaron energy $E(\mathbf{k})$, and $Z(\mathbf{k})$-factor at momentum $\mathbf{k}$ can be extracted from the Green's function dependence on imaginary time $\tau$. In the asymptotic limit $\tau \to \infty$, this dependence is governed by the ground state in the corresponding momentum sector, as follows from the spectral Lehman representation.
For the stable (non-decaying) quasiparticle state, we have

\begin{equation}
G(\mathbf{k},\tau \to \infty) \rightarrow Z(\mathbf{k}) e^{-[E(\mathbf{k})-\mu] \tau} \, .
\label{limit}
\end{equation}
%In the absence of additional stable quasiparticle states, the spectral density is zero up to the threshold, $E_{\rm th} = E(\mathbf{k})+ \omega_{\rm ph} $, for emission of the optical phonon. Thus, the leading finite-$\tau$ correction to Eq.~(\ref{limit}) starts with an additional exponential factor $e^{-\omega_{\rm ph} \tau}$. 
The $Z$ factor $Z(\mathbf{k})=|\langle \mathbf{k} | \tilde{\mathbf{k}} \rangle| ^2$ is given by the overlap between the polaron eigenstate $| \tilde{ \mathbf{k}} \rangle$ and the free-electron state $|\mathbf{k}\rangle=c^{\dagger}_{\mathbf{k}} |0\rangle$. The effective polaron mass are obtained from $m^*/m_0=2t/\frac{\partial ^2 E(\mathbf{k})}{\partial \mathbf{k}^2}$, where the bare electron mass is $m_0=1/2a^2t$, with the lattice spacing $a=1$. The chemical potential $\mu$ here is added to shift energies and control the rate of the exponential decays. It is used for computational convenience.

%In the deep adiabatic regime, the effective mass as a function of electron-phonon coupling strength $g/t$ changes its slope, which indicting there exists a crossover (see the discussion later). The nature of the crossover could be a transformation of one polaron state or a hybridization between two polaron states: light and heavy. In order to know the natural of the crossover, we include the correction part in the Green's function. The corresponding correction is included in fitting the data at large enough times:
%\begin{equation}
%G(\mathbf{k},\tau \gg \omega_{\rm ph}^{-1}) \, \rightarrow \, Z(\mathbf{k}) e^{-[E(\mathbf{k})-\mu ] \tau}  [1 + C e^{-K \tau} ] \, .
%\label{Gfit}
%\end{equation}
%If $K=\omega_0$, this means the polaron is in a stable quasiparticle state. As approaching the crossover, if $K$ changes value which is significantly smaller than $\omega_0$, it means there exists a crossover from a light polaron state to a heavy polaron state. 

The newly developed Quantum Monte Carlo method based on the path-integral formulation of the particle sector in combination with the real space diagrammatic of the phonon sector is used here to study the effects of a finite dispersion of the optical phonon mode on the properties of the bond polaron~\cite{PhysRevB.105.L020501}, especially the effective mass. We explore a large adiabatic regime with $\omega_0/t$ as low as 0.3 and phonon bandwidth as large as $W=1.5\omega_0$.

\section{Results and discussion}
\label{sec:sec4}

In this section, a comprehensive study of the bond polaron's properties as a function of electron-phonon coupling $g/t$ for different phonon frequency $\omega_0/t$ in the deep adiabatic regime and different phonon bandwidth $W/t$ (smaller, equal, and larger than the phonon frequency) are provided. Although the properties of the bond polaron at $\omega_0/t=0.5$ was investigated in Ref~\cite{PhysRevB.104.035143} using the Diagrammatic Monte Carlo method in the momentum space, we also put the results here for self-consistency and to set the stage of the discussion of the results with positive phonon bandwidth. We verify the correctness of our results by recalculating the phonon properties at $\omega_0/t=0.5$ using the newly developed Quantum Monte Carlo method~\cite{PhysRevB.105.L020501}.

\begin{figure}[t]
\includegraphics[width=0.41\textwidth]{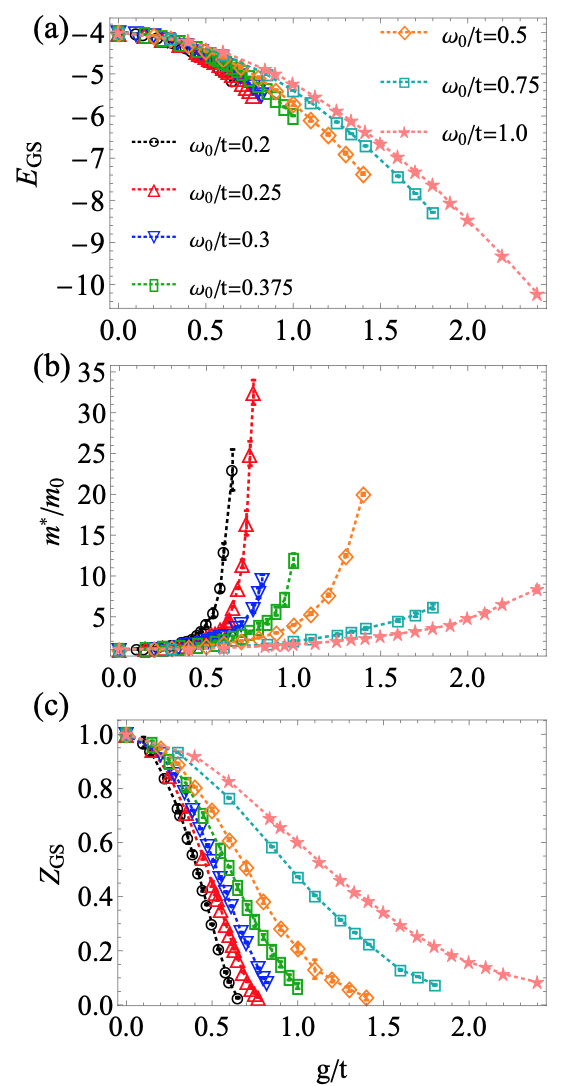}
\caption{The properties of the bond polaron: ground state energy $E_{GS}$ (a), effective mass $m^*/m_0$ (b), and $Z$ factor (c) as a function of electron-phonon coupling $g/ t$ for dispersionless phonon frequency $\omega_0/t=0.2$ (black circles), 0.25 (red up triangles), 0.3 (blue down triangles), 0.375 (green rectangles), 0.5 (orange diamonds), 0.75 (cyan squares), and 1.0 (pink stars). If not visible, error bars are within symbol size.
} 
\label{FIG1}
\end{figure}

\textit{Dispersionless case:}
%Figure~\ref{FIG1} shows the ground state energy $E_{GS}$ (a), the effective mass $m^*/m_0$ (b), and the $Z$ factor (c) of the dispersionless bond polaron as a function of electron-phonon coupling $g/t$ with different phonon frequencies varies from $\omega_0/t=1.0$, 0.75, 0.5, 0.375, 0.3, 0.25 to $0.2$ (in deep adiabatic regime). The $E_{GS}(g/t)$ curves for all the phonon frequencies indicate that the ground state evolves smoothly with $g/t$. At the same electron-phonon coupling $g/t$, the ground state energy decreases fast as the system goes into the deep adiabatic regime. The effective mass increases as the phonon frequency decrease for the same electron-phonon coupling $g/t$. At $\omega_0/t=1.0$, the effective mass increases linearly at weak and medium $g/t \sim 2.0$, then tends to change the slope. This trend becomes more clear as lowering the phonon frequency $\omega_0/t$. When $\omega_0/t=0.2$, the effective mass increases until coupling $g/t$ around 0.4 and becomes exponential. This indicts there exists a crossover and the nature of the crossover is the transformation from a light polaron state to a heavy polaron state (see the discussion later) in the deep adiabatic regime. The quasi-particle residue $Z$ decreases smoothly as a function of $g/t$ but tends to drop to zero quickly for smaller $\omega_0/t$. The effect of lowering the phonon frequency $\omega_0/t$ resulting a much heavy polaron in the lower coupling strength $g/t$.
Figure~\ref{FIG1} illustrates the main findings of our study for the dispersionless bond polaron in the deep adiabatic regime. We present the ground state energy $E_{GS}$ (a), the effective mass $m^*/m_0$ (b), and the $Z$ factor (c) as functions of the electron-phonon coupling $g/t$, while considering different phonon frequencies, ranging from $\omega_0/t=1.0$, 0.75, 0.5, 0.375, 0.3, 0.25 to $0.2$. The $E_{GS}(g/t)$ curves for all phonon frequencies demonstrate a smooth evolution of the ground state energy with increasing $g/t$. Remarkably, the ground state energy decreases rapidly as the system enters the deep adiabatic regime. Regarding the effective mass, we observe a consistent increase as the phonon frequency decreases, for a fixed electron-phonon coupling $g/t$. Specifically, at $\omega_0/t=1.0$, the effective mass exhibits a linear growth at weak and medium coupling strengths ($g/t \sim 2.0$) before experiencing a change in slope. This trend becomes more pronounced with decreasing $\omega_0/t$. When $\omega_0/t=0.2$, the effective mass shows a linear increase until the coupling strength reaches around $g/t\approx 0.4$, beyond which it grows exponentially. This behavior signifies the existence of a crossover, wherein the polaron undergoes a change from a light to a heavy state in the deep adiabatic regime, as discussed in detail later. Furthermore, the quasi-particle residue $Z$ exhibits a smooth decrease as a function of $g/t$, with a rapid drop observed for smaller $\omega_0/t$. Consequently, lowering the phonon frequency $\omega_0/t$ leads to a significantly heavier polaron at lower coupling strengths $g/t$.

\begin{figure}[t]
\includegraphics[width=0.43\textwidth]{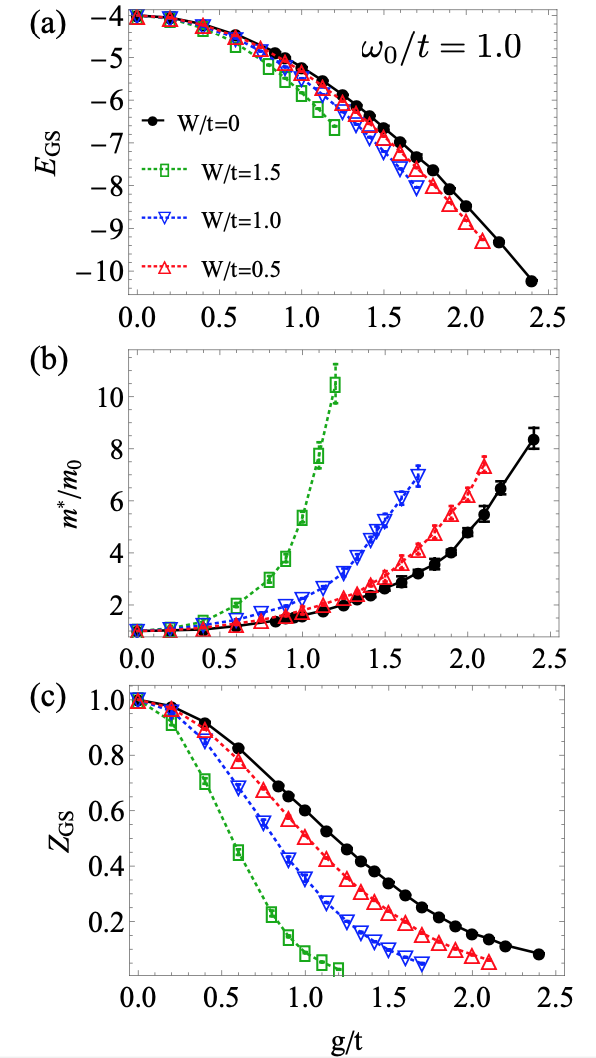}
\caption{The properties of the bond polaron with dispersive phonons: ground state energy $E_{GS}$ (a), effective mass $m^*/m_0$ (b), and $Z$ factor (c) as a function of the coupling strength $g/t$ for dispersionless phonon frequency $\omega_0/t=1.0$ (black circles) and dispersive case with $\omega_0/t=1.0$ and phonon bandwidth $W/t=0.5$ (red up triangles), $W/t=1.0$ (blue down triangles), and $W/t=1.5$ (green rectangles). If not visible, error bars are within the symbol size.
}
\label{FIG2}
\end{figure}

\textit{Dispersive case:} Figure~\ref{FIG2} shows the ground state energy $E_{GS}$ (a), the effective mass (b), and the $Z$ factor (c) for phonon bandwidth $W/t=0.5$, 1.0, and 1.5 (representing the bandwidth smaller, equal, and larger compared to the phonon frequency) in the adiabatic regime $\omega_0/t=1.0$. In comparison to the dispersionless case with a phonon frequency of $\omega_0/t=1.0$, the ground state energy $E_{GS}$ decreases smoothly as the coupling strength increases, for all three bandwidths. Additionally, at the same electron-phonon coupling strength, the ground state energy is lower for a larger phonon bandwidth.

The effective mass $m^*/m_0$ increases as a function of coupling strength $g/t$ and changes its slope for larger coupling strength for all three phonon bandwidths $W/t$. However, due to the phonon dispersion, the phonon frequency is lowered as $\omega_L=\omega_0-W/2$. For instance, when comparing the phonon bandwidth $W/t=1.5$, the phonon frequency is lowered to $\omega_L/t=0.25$. At a coupling strength of around $g/t \sim 0.75$, the effective mass $m^*/m_0$ is approximately 2.6. In contrast, for the dispersionless case $\omega_0/t=0.25$ (red upper triangular in Fig.~\ref{FIG1}(b)), the effective mass increases exponentially around $g/t \sim 0.6$ and is approximately 26.0 for $g/t \sim 0.75$. Compared to the dispersionless case, the effective mass with dispersive phonons is lighter. When comparing with the lower frequency $\omega_0=\omega_L$, with the positive phonon dispersion, we tend to have a light polaron at strong coupling strength. The effective mass increases linearly at a weak coupling strength and becomes exponential at a strong coupling strength for a large phonon bandwidth $W/t=1.5$. This indicates that there exists a crossover from a light polaron to a heavy polaron state (see the discussion later).

The quasiparticle residue decreases smoothly as a function of $g/t$ for all three bandwidths. At the same electron-phonon coupling $g/t$, the $Z$ factor drops rapidly to zero as the bandwidth $W/t$ increases.

\begin{figure}[h]
\includegraphics[width=0.425\textwidth]{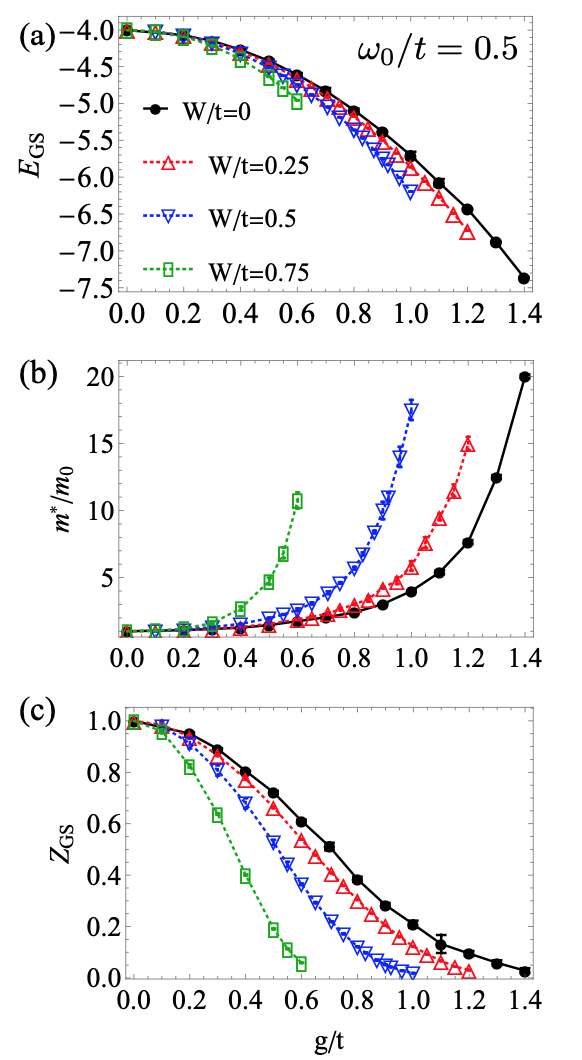}
\caption{The properties of the bond polaron with dispersive phonons: ground state energy $E_{GS}$ (a), effective mass $m^*/m_0$ (b), and $Z$ factor (c) as a function of the coupling strength $g/t$ for dispersionless phonon frequency $\omega_0/t=0.5$ (black circles) and dispersive case with $\omega_0/t=0.5$ and phonon bandwidth $W/t=0.25$ (red up triangles), $W/t=0.5$ (blue down triangles), and $W/t=0.75$ (green rectangles). If not visible, error bars are within the symbol size.
} 
\label{FIG3}
\end{figure}

\begin{figure}[h]
\includegraphics[width=0.42\textwidth]{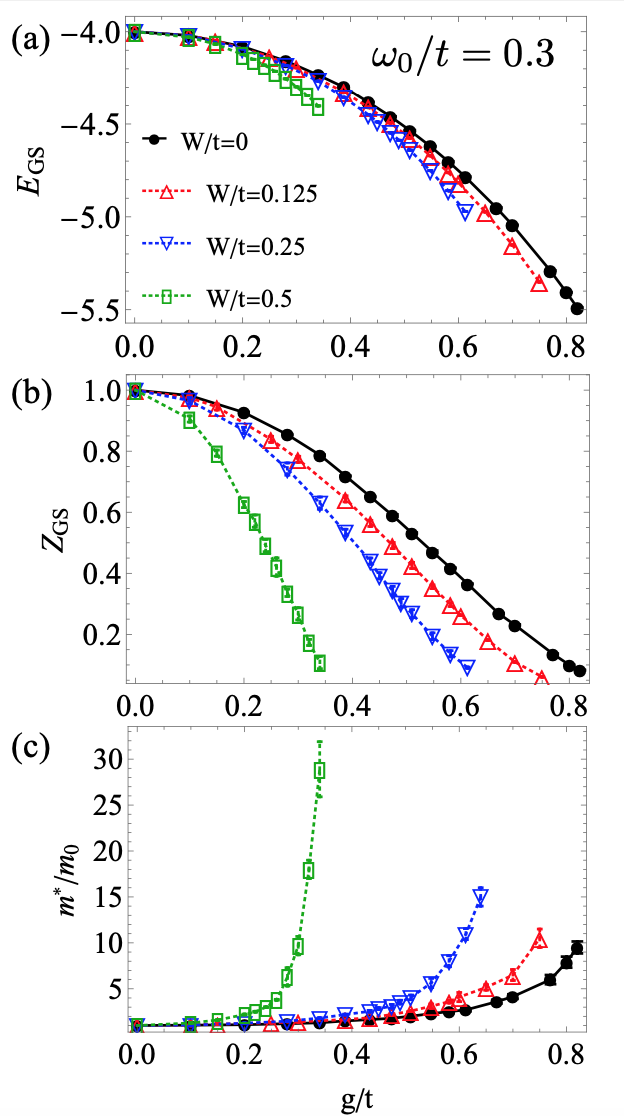}
\caption{The properties of the bond polaron with dispersive phonons: ground state energy $E_{GS}$ (a), effective mass $m^*/m_0$ (b), and $Z$ factor (c) as a function of the coupling strength $g/t$ for dispersionless phonon frequency $\omega_0/t=0.3$ (black circles) and dispersive case with $\omega_0/t=0.3$ and phonon bandwidth $W/t=0.125$ (red up triangles), $W/t=0.25$ (blue down triangles), and $W/t=0.5$ (green rectangles). If not visible, error bars are within the symbol size.
} 
\label{FIG4}
\end{figure}

The trend of the polaron properties is the same as the system goes into deep adiabatic regime $\omega_0/t=0.5$ for phonon bandwidth $W/t=0.25$, 0.5, and 0.75 (shown in Fig.~\ref{FIG3}) and $\omega/t=0.3$ for phonon bandwidth $W/t=0.125$, 0.25, and 0.5 (shown in Fig.~\ref{FIG4}). However, the change of the effective mass becomes more abrupt. 
%The phonon frequency of $\omega_0/t=0.5$ and bandwidth $W/t=0.5$ is lower to $\omega_L/t=0.25$. The effective mass starts to grow exponentially around $g/t=0.5$. Compare this to the dispersionless case with $\omega_0/t=0.25$ (red up triangular in Fig.~\ref{FIG1}) where the exponential growth of effective mass starts around $g/t \sim 0.55$ and $m^*/m_0 \sim 26.0$ at $g/t \sim 0.75$, the effective mass is around $4.6$ at $g/t\sim 0.75$ which is much lighter in the dispersion case. 
For instance, at $\omega_0/t=0.5$ with bandwidth $W/t=0.5$, the phonon frequency is lowered to $\omega_L/t=0.25$, and the effective mass starts to increase exponentially around $g/t=0.5$. When compared to the dispersionless case with $\omega_0/t=0.25$ (red upper triangular in Fig.~\ref{FIG1}(b)), where the exponential growth of the effective mass starts around $g/t \sim 0.6$ and reaches $m^*/m_0 \sim 26.0$ at $g/t \sim 0.75$, the effective mass is around $4.6$ at $g/t\sim 0.75$ in the dispersion case. Thus, the polaron in the dispersive case is much lighter. 

Similarly, for $\omega_0/t=0.3$ with bandwidth $W/t=0.25$, the phonon frequency is lowered to $\omega_L/t=0.05$. The effective mass starts to increase exponentially around $g/t \sim 0.22$. Although we cannot obtain the exact value of the effective mass for $\omega_0/t=0.05$ in the dispersionless case due to numerical challenges, it should be much heavier compared to the dispersive case, based on the trends shown in Fig.~\ref{FIG1} (b). Furthermore, at the same phonon frequency $\omega_0/t$ and electron-phonon coupling strength $g/t$, the effective mass increases more rapidly as the bandwidth increases. This observation can be explained by the fact that in the dispersive case, phonons are more mobile, leading to a more extended phonon cloud. Consequently, this extended phonon cloud effectively increases the effective mass of the electron by creating more obstacles for it to overcome as it moves through the lattice.

Overall, the results from $\omega_0/t=1.0$, 0.5, and 0.3 in the presence of dispersive phonons suggest the formation of relatively lighter polarons compared to the dispersionless case with the phonon frequency $\omega_0=\omega_L$.

%For $\omega_0/t=0.3$ with bandwidth $W/t=0.25$, the phonon frequency is lowered to $\omega_L/t=0.05$. The effective mass starts to increase exponentially around $g/t \sim 0.22$. Although we could not obtain the value of the effective mass for $\omega_0/t=0.05$ in the dispersionless case since it is numerically hard, the effective mass should be much heavier compared to the dispersive case due to the trend shown in Fig.~\ref{FIG1} (b). At the same phonon frequency $\omega_0/t$ and electron-phonon coupling strength $g/t$, the effective mass increases rapidly as the bandwidth increases. This can be explained as follows: in the dispersive case, phonons are more mobile, resulting in a more extended phonon cloud. This extended phonon cloud can effectively increase the mass of the electron by creating more obstacles for it to overcome as it moves through the lattice.

%From $\omega_0/t=1.0$, 0.5 and 0.3 results for the dispersive phonons, we can get a relatively light polaron in the dispersive case with lowered phonon frequency $\omega_L=\omega_0-W/2$ compared to the dispersionless case with bare phonon frequency $\omega_0=\omega_L$. 

\begin{figure}[h]
\includegraphics[width=0.5\textwidth]{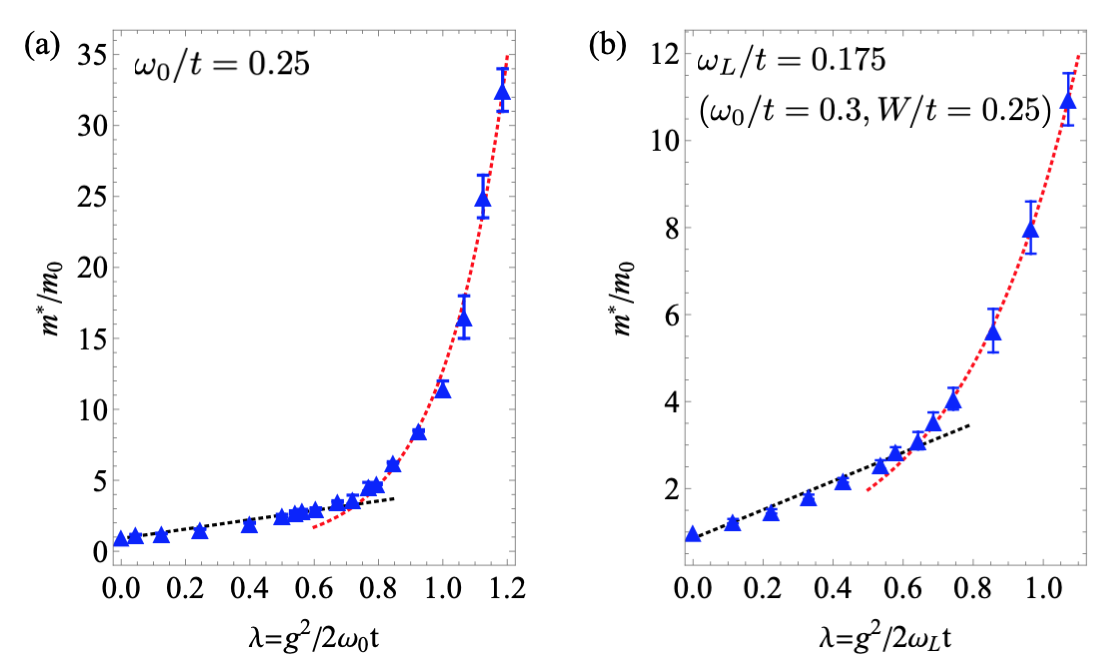}
\caption{(a) The effective mass $m^*/m_0$ as a function of the effective coupling $\lambda=g^2/2\omega_0 t$ at $\omega_0/t=0.25$. (b) The effective mass as a function of the effective coupling $\lambda=g^2/2\omega_L t$ at $\omega_0/t=0.3$
and phonon bandwidth $W/t=0.25$. The black dotted line is a linear fit and the red dotted line is an exponential fit for the effective mass as a function of effective coupling $\lambda$. There exists a crossover from a light bond polaron to an exponentially increasing heavy polaron as the electron-phonon coupling $g/t$ increases.
} 
\label{FIG5}
\end{figure}

%As we mentioned before, there exists a crossover from a light polaron state to a heavy polaron state in the deep adiabatic regime for both the dispersionless and dispersion case. The nature of the crossover is further investigated as shown in Fig.~\ref{FIG5}, where we study the effective mass as a function of coupling strength $\lambda$ for (a) dispersionless case with $\omega_0/t=0.25$ and (b) dispersive case with $\omega_0/t=0.3$ and phonon bandwidth $W/t=0.25$, in this case the phonon frequency is lowered to $\omega_L/t=0.175$. The black dotted line is a linear fit and the red dotted line is an exponential fit for the effective mass as a function of coupling strength $\lambda$ defined in Eq.~\ref{lambda}. There exists a crossover from a light polaron state to an exponentially increasing heavy polaron state at $\lambda \sim 0.75$ for the dispersionless case $\omega_0/t=0.25$ and $\lambda \sim 0.62$ for the dispersive case with $\omega_L/t=0.175$. %The nature of the crossover is distinguished by analyzing the Green's function as discussed in Section~\ref{sec:sec3}.  

As previously mentioned, a crossover from a light polaron state to a heavy polaron state in the deep adiabatic regime occurs for both the dispersionless and dispersive cases. We further investigate the nature of this crossover in Fig.~\ref{FIG5}, where we study the effective mass as a function of the effective coupling $\lambda$ for (a) the dispersionless case with $\omega_0/t=0.25$ and (b) the dispersive case with $\omega_0/t=0.3$ and phonon bandwidth $W/t=0.25$. In the dispersive case, the phonon frequency is lowered to $\omega_L/t=0.175$. In Fig.\ref{FIG5}, the black dotted line represents a linear fit, and the red dotted line represents an exponential fit for the effective mass as a function of the effective coupling $\lambda$ defined in Eq.\ref{lambda}. We observe that there exists a crossover from a light polaron state to a heavy polaron state with an exponential increasing of effective mass at $\lambda \sim 0.75$ for the dispersionless case with $\omega_0/t=0.25$ and at $\lambda \sim 0.62$ for the dispersive case with $\omega_L/t=0.175$. The crossover is characterized by an abrupt change in slope of the effective mass, showing a significant difference in the polaron's nature at the critical effective coupling in the deep adiabatic regime in both dispersionless and dispersive cases.

\section{Conclusion}
\label{sec:sec5}

%We have employed a recently developed Quantum Monte Carlo method, which utilizes the path-integral formulation of the particle sector along with the real-space diagrammatic approach of the phonon sector, to investigate the impact of finite dispersion of the optical phonon mode on the properties of the bond polaron. Specifically, we have analyzed the ground state energy, effective mass, and $Z$ factor of the bond polaron, considering different phonon frequencies as low as $\omega_0/t=0.3$, which lies in the deeply adiabatic regime, as well as the effect of positive phonon bandwidth on the bond polaron properties. Our findings reveal that in the absence of dispersion, effective mass increases as the phonon frequency decreases, provided the electron-phonon coupling is kept constant. Compared to the dispersionless case with $\omega_0=\omega_L$, the dispersive case exhibits lighter effective mass, especially at a strong electron-phonon coupling strength. Furthermore, our investigations demonstrate that in the deeply adiabatic regime, there is a transition from a linear increase in effective mass to an exponential increase in effective mass for both the dispersionless and dispersive cases. This transition corresponds to the transformation from a light polaron state to a heavy polaron state. The lighter effective mass for the dispersive case at strong coupling suggests the possibility of achieving a compact and light bi-polaron in the strong coupling limit, thereby increasing $T_c$ for the bond model.

We employed a recently developed Quantum Monte Carlo method, which combines the path-integral formulation of the particle sector with the real-space diagrammatic approach of the phonon sector, to investigate the effects of finite dispersion in the optical phonon mode on the properties of the bond polaron. Our analysis focused on the ground state energy, effective mass, and $Z$ factor of the bond polaron, considering different phonon frequencies as low as $\omega_0/t=0.3$, corresponding to the deeply adiabatic regime, as well as the impact of positive phonon bandwidth on the bond polaron's properties.

Our findings revealed that, in the absence of dispersion, the effective mass increases as the phonon frequency decreases while keeping the electron-phonon coupling constant. In comparison, the dispersive case exhibited lighter effective masses, particularly at strong electron-phonon coupling strengths.

Moreover, our investigations demonstrated a crossover from a linear increase in effective mass to an exponential increase in effective mass in the deeply adiabatic regime, both for the dispersionless and dispersive cases. Notably, the dispersive case showed a lighter effective mass at strong coupling, suggesting the possibility of achieving a compact and light bipolaron in the strong coupling limit, potentially leading to an increase in the critical temperature $T_c$ for the bond model. Achieving the higher $T_c$ in the bond SSH model requires that the relative materials to be in the same parameter regime as describe in Ref~\cite{PhysRevX.13.011010,Johnarxiv2022}, which means (i) the interference of tunneling pathways controlled by light atoms, for example, hydrogen with large $\omega_0$, (ii) the tunneling amplitude should be relatively small, not significantly larger than the bottom of the phonon dispersion, (iii) large dielectric constant to reduce the destructive effect of long-range Coulomb interaction. With a possible compact and light bipolaron at strong coupling limit with dispersive phonons, there will be more possibilities in the search for new materials with high superconducting temperatures.

\begin{acknowledgments}
C.Z. thanks Nikolay Prokof’ev and Boris Svistunov for helpful discussions. This work is supported by the National Natural Science Foundation of China (NSFC) under Grants No. 12204173 and No. 12275263), the Innovation Program for Quantum Science and Technology (under Grant No. 2021ZD0301900), and the National Key R $\&$ D Program of China (under Grant No. 2018YFA0306501).

\end{acknowledgments}

\bibliography{dispersive}

\end{document}